\documentclass[12pt]{article}
\usepackage{amsmath, amsthm, amssymb}
\usepackage{graphicx}

\begin{document}

\title{Correlations of photon trajectories in the problem of light scintillations}

\author{R. A. Baskov \, and O. O. Chumak\footnote{Corresponding author: chumak@iop.kiev.ua} \\[3mm]
 \\[5mm] Institute of Physics of the National Academy of Sciences\\ pr. Nauki 46, Kyiv-28, MSP 03028 Ukraine \\[3mm]\rule{0cm}{1pt}}

\maketitle %\markright{right_head}{LA-UR-xx-xxxxx}

\begin{abstract}

A distribution function approach is applied to describe the dynamics
of the laser beam in the Earth atmosphere. Using a formal solution
of the kinetic equation for the distribution function, we have
developed an iterative scheme for calculation of the scintillation
index ($\sigma^2$). The problem reduces to obtaining the photon
trajectories and their correlations. Bringing together theoretical
calculations and many-fold computer integrations, the value of
$\sigma^2$ is obtained. It is shown that a considerable growth of
$\sigma^2$ in the range of a moderate turbulence is due to the
correlations of different trajectories. The criteria of
applicability of our approach for both the coherent and partially
coherent light are derived.

\end{abstract}

\section{Introduction}

Basic principles of the radiation transfer theory were formulated in
the seminal paper of Schuster \cite{Schuster} as early as the
beginning of the $20$th century. The paper \cite{Schuster} was
devoted to the light propagation in a foggy atmosphere. Since then
the Schuster approach has obtained many applications in such
important fields as astronomy, laser communication and radar
systems, remote sensing etc.

The range and performance of light communication systems are limited
significantly by an unfavorable influence of local fluctuations of
the refractive index of the Earth atmosphere. On the other hand,
high sensitivity of the photon trajectories to the fluctuations can
be used for the atmosphere diagnostics \cite{mil}. The key issue
about the index-of-refraction structure constant $C_n^2$ is that it
cannot be reliably computed from first principles. The
refractive-index fluctuations arise from temperature inhomogeneities
of the air. The inhomogeneities cause turbulent eddies which give
rise to a random distribution of the air
density(\cite{Tatarskii1}-\cite{Andrews1}). This results in random
spatial variations of the refractive index.

The turbulent eddies are described by a wide range of characteristic
lengths of inhomogeneities. These lengths cover the interval from
few millimeters (the inner radius, $l_0$) to hundred meters (the
outer radius, $L_0$). Therefore, various types of beam scattering
are observed. The scattering by large-size eddies results in random
redirections of the beam as a whole. This process is known in the
literature as a "wandering" or "dancing" of the beam
\cite{Fante1},\cite{xia}. On the other hand, the scattering by
small-size eddies causes spreading of the beam. For a long-distance
propagation or a strong turbulence, the beam radius becomes greater
than the characteristic sizes of the inhomogeneities. In this case
the probability of the beam to be redirected becomes small and the
relative value of the wandering radius decreases \cite{Chumak
wander}.

The beam wandering and broadening can be considered as the specific
manifestations of a more general phenomenon, namely the intensity
fluctuations (i.e. scintillations) caused by the atmosphere
turbulence. The scintillations have a tendency of saturating for a
long-distance
 propagation \cite{gra},\cite{fan} (the regime of a strong turbulence).
This is because in the course of propagation the radiation acquires
the properties of the Gaussian statistics when the signal-to-noise
ratio (SNR) tends to unity. The asymptotic behavior of the
scintillation index, $\sigma^2\rightarrow 1$, was explained in Refs.
\cite{Wang}-\cite{Lee}. Moreover, it was shown quite generally that
this property stays unchanged for any refractive index distribution,
provided the response time of the recording instrument is short
compared with the source coherence time. This result was confirmed
analytically in  \cite{Ban55}.

At the same time, calculations, performed by different methods in
\cite{Ban54} and \cite{Chu}, show a possibility of significant
suppression of the scintillations. To this end partially coherent
laser beams with the coherence time shorter than the detector
integration time (a slow detector) can be used. The case of a
partial coherence was also studied in \cite{cor},\cite{coro}. Recent
theoretical and experimental developments on propagation of
partially coherent beams in a turbulent atmosphere were discussed in
\cite{fei}.

There are several analytical approaches explaining behavior of the
scintillation index in the case of strong turbulence
\cite{Chu,Das,Yak}. Their analysis is based on the physical picture
where four waves, forming the second moment of the intensity,
conserve only pair correlations in course of long-distance
propagation. Two different pairs of the photon trajectories
contribute into the square of the photon density at the detector.
Dashen used the Feynman path integrals to prove that in a convincing
manner \cite{Das}.

The  recent interest to beam propagation was awakened  by the
development of quantum communication in the free atmosphere
\cite{Fed}, \cite{Hug}. Detailed studies of the effect of the
turbulence-induced losses on the quantum state of the light in the
course of satellite-mediated communication  and for realization of
the entanglement transfer in the atmosphere were reported in Refs.
\cite{Bou} and \cite{Sem}.

The formalism of the photon distribution function (the photon
density in the coordinate-momentum space \cite{UJP}), is also
applicable to the problem of scintillations \cite{Chumak
wander,Chu,ChuSingle,ChuPhase}. The mentioned papers are based on a
physical picture, which is similar to the described above. The
method of photon distribution function is used for  description of
both the classical and the quantum light including propagation of
single-photon pulses (see, for example, Refs.
\cite{ChuSingle,sto,sto14}). Solution of the kinetic equation for
the operator of photon density is based on the method of
characteristics. The assumption of weak disturbances of photon
momenta by the  atmosphere (the paraxial approximation) reduces the
problems of scintillations to the problem of obtaining photon
trajectories and their correlations. A slowly varying fluctuating
force, deflecting photon trajectories from straight lines, describes
the effect of the atmospheric eddies.

In this work we study the scintillation index for moderate and
strong turbulences, when correlation of trajectories of only two
photons is required. Accuracy of the calculations depends on the
accuracy of obtaining the trajectories. Using high-order iterations
and bringing together analytical and numerical procedures, we
calculate the scintillation index. Our main interest is to analyze
the range of moderate turbulence strengths where previous theories
do not ensure a reliable description. Comparison of the obtained
results with those represented in \cite{Chu} helps indicate the
range of turbulence where a simplified approach should be corrected
by high-order iterations. Also, our studies describe more
realistically the effect of partial coherence.

\section{Photon distribution function approach}

The photon distribution function is defined by analogy with
distribution functions in solid state physics. In particular, it is
similar to the phonon distribution function. Both of them are
defined as \cite{UJP},\cite{chuZ}
\begin{equation}\label{threee}
f({\bf r},{\bf q},t)=\frac 1V\sum_{\bf k}e^{-i{\bf kr}}b^\dag_{{\bf
q}+ {\bf k}/2}b_{{\bf q}-{\bf k}/2},
\end{equation}
where $b^\dag_{\bf q}$ and $b_{\bf q}$ are the bosonic creation and
annihilation operators of photons or phonons with the momenta ${\bf
q}$, and $V\equiv L_xL_yL_z$ is the normalizing volume. Polarization
of the corresponding modes is not specified in (\ref{threee}). In
the paraxial approximation, assumed here, the initial polarization
of the beam remains almost unchanged even for a long-distance
propagation (see, for example, Ref. \cite{stroh}).

The operator $f({\bf r},{\bf q},t))$ describes the photon (phonon)
density in the phase ({\bf r,q}) space. Usually, the characteristic
sizes of spatial inhomogeneities of the radiation field are much
greater than the wave-length. In this case the sum in Eq.
(\ref{three}) can be restricted by small $k$. Here and in what
follows we consider that $k<k_0\ll q_0$, where $q_0$ is the wave
vector corresponding to the central frequency of the radiation,
$\omega _0=cq_0$. At the same time $k_0$ should be taken
sufficiently large to provide a required accuracy of the beam
profile description.

The evolution of the Heisenberg operator $f({\bf r},{\bf q},t)$ is
determined by the commutator
\begin{equation}\label{five}
\partial_t f({\bf r},{\bf q},t)=\frac 1{i\hbar }[f({\bf r},{\bf
q},t),H],
\end{equation}
where
\begin{equation}\label{six}
H=\sum_{\bf q}\hbar\omega_{\bf q}b^\dag_{\bf q}b_{\bf q}-\sum_{\bf
q,k}\hbar\omega_{\bf q}n_{\bf k}b^\dag_{\bf q}b_{\bf q+k},
\end{equation}
is the Hamiltonian of photons in a medium with a fluctuating
refractive index $n({\bf r})$ ($n_{\bf k}$ is its Fourie transform),
$\hbar\omega_{\bf q}=\hbar cq$ and ${\bf c_q}=\frac{\partial
\omega}{\partial{\bf q}}$ are the vacuum values of the photon energy
and velocity, respectively.

Assuming  the characteristic values of the photon momentum to be
much greater than the wave vectors of turbulence, the kinetic
equation for the photon distribution function can be written as
\begin{equation}\label{seven}
\{ \partial_t +{\bf c_q}\partial_{\bf r}+{\bf F}({\bf
r})\partial_{\bf q}\} f({\bf r},{\bf q},t)=0,
\end{equation}
where ${\bf F}({\bf r})= \omega _0\partial_{\bf r}n({\bf r})$ is the
random force originating from the atmospheric turbulence. The
general solution of Eq. (\ref{seven}) is given by
\begin{equation}\label{nine}
f({\bf r},{\bf q},t)=\phi \Bigg\{{\bf r}-\int _0^tdt^\prime\frac
{\partial {\bf r}(t^\prime)}{\partial t^\prime};{\bf
q}-\int_0^tdt^\prime\frac {\partial {\bf q}(t^{\prime})}{\partial
t^\prime}\Bigg\},
\end{equation}
where the function $\phi ({\bf r},{\bf q})$ is the "initial" value
of $f({\bf r},{\bf q},t)$, i.e.
\begin{equation}\label{ten}
\phi ({\bf r},{\bf q})=\frac 1V\sum_{\bf k}e^{-i{\bf kr}}(b^+_{{\bf
q}+ {\bf k}/2}b_{{\bf q}-{\bf k}/2})|_{t=0}\equiv \sum_{\bf
k}e^{-i{\bf kr}} \phi ({\bf k},{\bf q}).
\end{equation}
The derivatives $\frac {\partial{\bf r}(t^\prime)}{\partial
t^\prime}$ and $\frac {\partial{\bf p}(t^\prime)}{\partial
t^\prime}$ should satisfy the equations
\[ \frac {\partial {\bf r}(t^\prime)}{\partial t^\prime}={\bf c}[{\bf q}(t^\prime)] \]
\begin{equation}\label{eight}
\frac {\partial {\bf q}(t^\prime)}{\partial t^\prime}={\bf F}[{\bf
r}(t^\prime)],
\end{equation}
completed with the boundary conditions ${\bf r}(t^\prime )={\bf r}$
and ${\bf q}(t^\prime)={\bf q}$ for $t=t^\prime$. As we see, Eqs.
(\ref{eight}) coincide with the classical (the Newton) equations of
motion of a point particle moving with the velocity ${\bf c_q}$ and
affected by an external force ${\bf F}({\bf r})$. Formal solutions
of Eqs. (\ref{eight}) can be written as
\begin{equation}\label{eleven}
{\bf q}(t^\prime )={\bf q} +\int _t^{t^{\prime}} dt^{\prime \prime }
{\bf F}[{\bf r}({\bf q},t^{\prime \prime})]
\end{equation}
and
\begin{equation}\label{twelve}
{\bf r} (t^\prime )={\bf r}-{\bf c _q }(t-t^\prime) -\frac c{q_0}
\int _t^{t^{\prime }}dt^{\prime \prime }(t^{\prime \prime}-t^{\prime
}) {\bf F}[{\bf r}(t^{\prime \prime})],
\end{equation}
 Eqs. (\ref{eleven}) and (\ref{twelve}) allow us to rewrite the
 expression (\ref{nine}) as
 \begin{equation}\label{thirteen}
f({\bf r},{\bf q},t)=\phi \Bigg\{{\bf r}-{\bf c_q}t+\frac c{q_0}\int
_0^tdt^\prime t^\prime {\bf F} [{\bf r}({\bf q},t^\prime )];{\bf
q}-\int _0^tdt^\prime {\bf F} [{\bf r}({\bf q},t^\prime )]\Bigg\}.
\end{equation}
If ${\bf F}({\bf r})$ is a known function, an approximate value for
$f({\bf r},{\bf q},t)$ can be obtained by inserting  the term ${\bf
r}({\bf q},t^{\prime})\approx{\bf r}-{\bf c _q }(t-t^{\prime}$) into
Eq. (\ref{thirteen}). In this case the argument of the fluctuating
force ${\bf F}[{\bf r}({\bf q},t^{\prime})]$ is replaced by a
straight line, that is correct only in the absence of the
turbulence. Improvement of the theory can be achieved if the
argument of ${\bf F}$ accounts for the turbulence.

It follows from Eq. (\ref{thirteen}) that statistical properties of
the radiation depend not only on the turbulence but also on the
initial distribution function $\phi ({\bf r},{\bf q})$. This
function is determined by the source field. Its explicit form is
determined in the course of "sewing" of the near-aperture and the
atmospheric fields \cite{Chu} given by the amplitudes $b_{\bf q}
(b^\dag_{\bf q})$. We consider the light propagation in the
$z$-direction. The source field is assumed to be described by the
Gaussian function, $\Phi({\bf r})=(2/\pi)^{1/2}\frac
1{r_0}e^{-r^2_\bot/{r_0^2}}$. Then the propagating amplitudes are
given by
\begin{equation}\label{fourteen}
b_{{\bf q_\bot},q_0}(t=0)=b(2\pi/S)^{1/2}r_0e^{-q^2_\bot {r_0}^2/4},
\end{equation}
where $b$ is the near-aperture amplitude of the laser field, index
($_\bot$) means the perpendicular to the $z$-axis components, and
$S=L_xL_y$.

We will take into account the effect of the phase diffuser by
multiplying the distribution $\Phi({\bf r})$ by the phase factor
$e^{-i{\bf ar_\bot}}$  where the quantity ${\bf a}$ is a random
variable. In this case Eq. (\ref{fourteen}) should be modified by
substituting in its right-hand side  ${\bf q}_\bot+{\bf a}\equiv{\bf
q}_{\bf a}$ for ${\bf q_\bot} $. Such a simple modeling of the phase
diffuser is justified if (i) the detection time is much longer than
the characteristic time of the variation of ${\bf a}$ (slow
detector) and (ii) there is a large root mean square of the phase
fluctuations. (More detailed analysis is presented in
\cite{ChuPhase}.) This case corresponds to the Gaussian distribution
of ${\bf a}$:
\begin{equation}\label{fiftn}
P(a_{x,y})=\frac{\lambda}{2\pi^{1/2}}e^{-a_{x,y}^2{\lambda}^2/4},
\end{equation}
with a covariance $\langle a_{x,y}^2\rangle=\lambda ^{-2}$ and the
transverse correlation function of the outgoing field (at $t=0$) is
given by
\begin{equation}\label{sixstn}
\langle E({\bf r}_\bot)E({\bf r}_\bot+{\bf\Delta})\rangle_{\bf
a}=E^2_0e^{-[r^2_\bot+({\bf
r_\bot+\Delta})^2]r_0^{-2}}e^{-\Delta^2\lambda^{-2}}.
\end{equation}
Here $E_0=E(r_\bot=0,\Delta=0,t=0)$ and the notation $\langle
...\rangle_{\bf a}$ means averaging over distribution $P(a_{x,y})$.
The radiation, whose correlation properties are described by
function (\ref{sixstn}), is referred to as the Gaussian Shell-model
field. The parameter $\lambda$ in the exponential factor describes
the decrease of the transverse correlation length. It can also be
said that this parameter generates a new characteristic length,
$1/r_1$, in the momentum distribution (i.e., in the ${\bf
q}$-domain). This is seen from the explicit term for $\phi ({\bf
k},{\bf q})$ which after averaging over the fluctuations of ${\bf
a}$ reduces to
\begin{equation}\label{sevntn}
\langle\phi ({\bf k},{\bf q})\rangle_{\bf a}=2\pi\frac{b^\dag
b}{VS}r_1^2e^{-q^2_\bot \frac {r_1^2}2-k^2_\bot \frac {r_0^2}8},
\end{equation}
where $r_1^2=r_0^2\bigg(1+2r_0^2\lambda^{-2}\bigg)^{-1}$, and
variables $q_z$ and $k_z$ are omitted.

It is seen from Eq. (\ref{sevntn}) that $q_\bot$ is distributed in
the range of the order of $\sqrt 2/r_1$ that is greater than the one
for coherent beam. In contrast, the characteristic value of ${\tilde
k}$ depends only on the initial size of the beam ( ${\tilde
k}\sim\sqrt 8/r_0$).

In the course of light propagation, the diffraction phenomena and
scattering by atmospheric inhomogeneities broaden the beam resulting
in decrease of ${\tilde k}$. At the same time, the value of ${\tilde
q}$ increases with the distance. This is because of the
Brownian-like motion of photons in the ${\bf q_\bot}$-domain (see
Ref. \cite{Chu}). Such a simple physical picture, elucidating
evolution of the beam geometry, is, however, not applicable to the
description of scintillations. The phenomenon of scintillations is
more complecated and can be described in terms of spatio-temporal
correlations of four waves.

\section{Scintillation index}

The photon distribution function is used here to obtain the
scintillation index $\sigma^2$. The definition of $\sigma^2$ is
given by
\begin{equation}\label{eightn}
\sigma^2=\frac {\langle I^2({\bf r})\rangle-\langle I({\bf
r})\rangle ^2}{\langle I({\bf r})\rangle^2}.
\end{equation}
The photon density $I({\bf r},t)$ is expressed in terms of the
distribution function as
\begin{equation}\label{nintn}
I({\bf r},t)=\sum_{\bf q}f({\bf r},{\bf q},t)= 2\pi\frac{b^\dag
br_0^2}{SV}\sum_{\bf q,k}e^{-i{\bf k}[{\bf r}-{\bf c}({\bf q})t+
\frac c{q_0}\int _0^tdt^\prime t^\prime {\bf F}({\bf r}({\bf
q},t^\prime ))]- Q^2_{\bf a}\frac {r^2_0}2-k^2\frac {r^2_0}8},
\end{equation}
where ${\bf Q_a}\equiv {\bf Q+a}={\bf q+a}-\int _0^tdt^\prime {\bf
F}[{\bf r}({\bf q},t^\prime )$. The summation is taking over ${\bf
q_\bot}$ and ${\bf k_\bot}$ components, while $q_z$ and $k_z$ are
considered to be fixed: $q_z=q_0$ and $k_z=0$. The exponential term
originates from the solution (\ref{thirteen}) of the kinetic
equation (\ref{seven}).

To obtain  $\langle I({\bf r},t)\rangle$, three independent
averagings are required. One of them concerns the source variables.
In the case of a coherent state of the source, $|\beta\rangle$, we
have $\langle b^\dag b \rangle=|\beta|^2$. The second averaging over
a random phase of the diffuser should be carried out as explained by
Eq. (\ref{sevntn}). The third averaging deals with the fluctuating
force ${\bf F}$. These three actions can be performed independently
that facilitates the analysis. Also, the calculations are simplified
if we use the identity
\begin{equation}\label{twnt}
e^{-Q^2r^2_0/2}\equiv\int\frac {d\bf p}{2\pi r_0^2}e^{i{\bf
pQ}-p^2/2r_0^2}.
\end{equation}
Because of Eq. (\ref{twnt}), the term in the exponent of Eq.
(\ref{nintn}) reduces to the linear in ${\bf F}$ form. Then,
considering ${\bf F}$ as a random Gaussian variable, the value of
$\langle I({\bf r},t)\rangle$ can be easily obtained in a manner,
explained in Ref. \cite{Chu}. To calculate $\langle I({\bf
r},t)\rangle$, an explicit form of the refractive-index correlation
function, $\langle n({\bf r})n({\bf r}^\prime)\rangle$, is required.
In a statistically homogeneous atmosphere it can be written as
\begin{equation}\label{ad}
 \langle n({\bf r})n({\bf r}^\prime)\rangle=\int d{\bf g}
 e^{-i{\bf g}(\bf {r-r^\prime})}\psi({\bf g}).
\end{equation}
A widely used the  von Karman approximation for the spectrum,
$\psi({\bf g})$, is given by
\begin{equation}\label{add}
 \psi({\bf
 g})=0.033C_n^2\frac{\exp[-(gl_0/2\pi)^2]}{[g^2+L_0^{-2}]^{11/6}},\,
 |{\bf g}|\equiv g,
 \end{equation}
where the vector ${\bf g}$ is defined in the three dimensional
domain.

The "source" part of $\langle I^2({\bf r})\rangle$, given by
$\langle b^\dag b b^\dag b\rangle$, is approximately equal to
$\langle b^\dag b^\dag bb\rangle=|\beta|^4$, when the condition
$|\beta|^4>>|\beta|^2$ is satisfied. This inequality implies that
the initial laser radiation is in a multiphoton coherent state. The
averaging over independent random quantities ${\bf a}$ and ${\bf
a}^\prime$ can be used instead of the time averaging of the diffuser
state. Then we have
\[\langle I^2({\bf r},t)\rangle =\bigg|\frac {2\pi\beta^2
r_0^2}{VS}\bigg|^2 \sum_{{\bf q},{\bf k},{\bf q^\prime},{\bf
k^\prime}}\langle e^{-i{\bf k}[{\bf r}-{\bf c_q}t+\frac
c{q_0}\int_0^tdt^\prime t^\prime {\bf F}({\bf r_q}(t^\prime))]-i{\bf
k^\prime}[{\bf r}-{\bf c_{\bf q^\prime}} t+\frac
c{q_0}\int_0^tdt^\prime t^\prime {\bf F}({\bf
r_{q^\prime}}(t^\prime))]}\]
\begin{equation}\label{twnto}
 \times \big \{ e^{-(Q_{\bf a}^2+Q_{\bf a^\prime}^2+\frac
{k^2+{k^\prime} ^2}4 )\frac{r_0^2}2}+e^{-[({\bf Q_a}+\frac{\bf
k}2)^2+({\bf Q^\prime _a}-\frac{\bf k^\prime}2)^2+ ({\bf
Q_{a^\prime}}-\frac{\bf k}2)^2+({\bf Q^\prime_{a^\prime}}+\frac{\bf
k^\prime}2)^2]\frac{r_0^2}4}
 \big \}\rangle.
\end{equation}

There are two terms in the braces of Eq. (\ref{twnto}). They appear
only if the initial four-wave correlation reduces to the pair
correlation \cite{Chu}. Such a modification of the statistical
properties of the radiation occurs when the waves propagate for a
long time which is sufficient for randomization of the transverse
photon momentum. A more general case, which includes the regime of
fast detection, was analyzed in Ref. \cite{ChuPhase}.

The averaging of Eq. (\ref{twnto}) over ${\bf a}$ and ${\bf
a^\prime}$  results in
\[\langle I^2({\bf r},t )\rangle =\bigg|\frac {2\pi\beta^2
r_1^2}{VS}\bigg|^2 \sum_{{\bf q},{\bf k},{\bf q^\prime},{\bf
k^\prime}}\langle e^{-i\{{\bf k}[{\bf r}-{\bf c}({\bf q})t]+ {\bf
k}^\prime [{\bf r}-{\bf c}({\bf q}^\prime )t]+\frac c{q_0}\int
_0^tdt^\prime t^\prime [{\bf k}{\bf F}({\bf r}({\bf q},t^\prime
))+{\bf k}^\prime {\bf F}({\bf r}({\bf q}^\prime , t^\prime ))]\}}\]
\begin{equation}\label{twntt}
\times \big\{e^{-(Q^2+{Q^\prime }^2)r_1^2/2-(k^2+{k^\prime }^2)
r_0^2/8}+ e^{-[({\bf Q}-{\bf Q}^\prime )^2+ ({\bf k}+{\bf k}^\prime
)^2/4]r_0^2/4-[({\bf Q}+{\bf Q}^\prime )^2+ ({\bf k}-{\bf k}^\prime
)^2/4] r_1^2/4}\big \}\rangle.
\end{equation}
In the absence of a phase diffuser, $r_0=r_1$,  the summands in the
last braces contribute equally into (\ref{twntt}).

Similarly to Eq. (\ref{twnt}), the factor $e^{-(Q^2+{Q^\prime
}^2)\frac {r_1^2}2}$ in (\ref{twntt}) can be expressed in the
integral form as
\begin{equation}\label{twnth}
e^{-(Q^2+{Q^\prime }^2)\frac {r_1^2}2}=\int\frac{d{\bf p}d{\bf
p^\prime}}{(2\pi r_1^2)^2}e^{i{\bf pQ}+i{\bf p^\prime Q^\prime
}-(p^2+{p^\prime}^2 )/{2r_1^2}}.
\end{equation}
As we see, the exponent in the left-hand side is represented as a
linear form of the force  $\bf F$. A similar transform is applicable
to the second term in the last braces of (\ref{twntt}). As a result,
the fluctuating force enters the right-hand side of (\ref{twntt})
only via the common multiplier, $M$, given by
\begin{equation}\label{twntfo}
M=e^{-i\int_0^tdt^\prime \{ ({\bf p}+{\bf k}t^\prime c/q_0){\bf
F}[{\bf r}({\bf q},t^\prime )]+ ({\bf p}^\prime +{\bf k}^\prime
t^\prime c/q_0){\bf F}[{\bf r}({\bf q}^\prime,t^\prime )]\} }.
\end{equation}
Obtaining of the average value of $I^2$ reduces to averaging of $M$
with many-fold integration. Assuming the exponent in (\ref{twntfo})
as a Gaussian random variable, we can write
\begin{equation}\label{twntfi}
\langle M\rangle=e^{-\frac 12\big\langle\big(\int_0^tdt^\prime \{
({\bf p}+{\bf k}t^\prime c/q_0){\bf F}[{\bf r}({\bf q},t^\prime )]+
({\bf p}^\prime +{\bf k}^\prime t^\prime c/q_0){\bf F}[{\bf r}({\bf
q}^\prime,t^\prime )]\}\big)^2\big\rangle }\equiv
e^{-\frac12(\phi_{PP}+2\phi_{PP^\prime}+ \phi_{P^\prime P^\prime})}.
\end{equation}
Two types of the correlation functions determine $\langle M\rangle$:
\begin{equation}\label{twntsi}
\phi_{PP^\prime}=\int_0^t\int_0^tdt^\prime dt^{\prime\prime}  ({\bf
p}+{\bf k}t^\prime c/q_0)\cdot\langle{\bf F}[{\bf r}({\bf
q},t^\prime )] {\bf F}[{\bf r}({\bf q}^\prime,t^{\prime\prime}
)]\rangle\cdot({\bf p}^\prime +{\bf k}^\prime t^{\prime\prime}
c/q_0)] ,
\end{equation}
\begin{equation}\label{twntse}
\phi_{PP}=\int_0^t\int_0^tdt^\prime dt^{\prime\prime}  ({\bf p}+{\bf
k}t^\prime c/q_0)\cdot\langle{\bf F}[{\bf r}({\bf q},t^\prime )]
{\bf F}[{\bf r}({\bf q},t^{\prime\prime} )]\rangle\cdot({\bf p}
+{\bf k} t^{\prime\prime} c/q_0)],
\end{equation}
where symbols $P$ and $P^\prime$ denote sets of three vector
variables $P=\{{\bf q,p,k}\}$ and $P^\prime=\{{\bf
q^\prime,p^\prime,k^\prime}\}$.
 The correlation functions of the forces
along different  ({$\bf q\neq\bf q^\prime$}) and coinsiding ($\bf
q=\bf q^\prime$) trajectories  enter Eqs. (\ref{twntsi}) and
(\ref{twntse}), respectively. The former can be rewritten as
\begin{equation}\label{twntei}
\langle F_\alpha[{\bf r}({\bf q},t^\prime )]  F_\beta[{\bf r}({\bf
q}^\prime,t^{\prime\prime} )]\rangle=\langle{F_\alpha}[{\bf r}({\bf
q},t^\prime )-{\bf r}({\bf q^\prime},t^{\prime\prime })]
{F_\beta}[0]\rangle,
\end{equation}
where the notations $\alpha$ and $\beta$ stand for the $x$ and $y$ -
components. The expression for (\ref{twntse}) follows from Eq.
(\ref{twntei}) by setting ${\bf q=q}^\prime$.

The right-hand side of Eq. (\ref{twntei}) is assumed to be a
function of the coordinate difference, ${\bf r}({\bf q},t^\prime
)-{\bf r}({\bf q^\prime},t^{\prime\prime})$. It is so if the
atmosphere is statistically homogeneous. In the course of averaging,
dependence of the coordinate difference on the fluctuating force
should be also taken into account. This dependence is given by the
relation
\[{\bf r}({\bf q},t^\prime )-{\bf r}({\bf
q^\prime},t^{\prime\prime})=({\bf e}_zc+{\bf
c_{q^\prime}})(t^\prime-t^{\prime\prime})-{\bf
c_{q-q^\prime}}(t-t^{\prime})+\frac
c{q_0}\int_{t^\prime}^{t^{\prime\prime}}dt_1 (t^\prime -t_1){\bf
F}[{\bf r}({\bf q}^\prime,t_1)]\]
\begin{equation}\label{twntn}
+\frac c{q_0}\int_t^{t^\prime}dt_1 (t^\prime -t_1)\{{\bf F}[{\bf
r}({\bf q},t_1)]-{\bf F}[{\bf r}({\bf q}^\prime,t_1)]\},
\end{equation}
which follows from Eq. (\ref{twelve}). The distance $|{\bf r}({\bf
q},t^\prime )-{\bf r}({\bf q^\prime},t^{\prime\prime})|$ should be
of the order or less than the outer radius, $L_0$, of the
turbulence. Taking into account that $c>>|c_{\bf q-q^\prime}|,
|c_{\bf q^\prime}|$, we infer that $|t^\prime-t^{\prime\prime}|\leq
L_0/c$. This means that in the right-hand side of Eq. (\ref{twntn})
$c_{\bf q^\prime}$ in the first term  and the third term, which is
proportional to $(t^\prime-t^{\prime\prime})^2$, can be omitted.
Then Eq. (\ref{twntn}) reduces to
\[ {\bf r}({\bf q},t^\prime )-{\bf r}({\bf
q^\prime},t^{\prime\prime})={\bf
e}_zc(t^\prime-t^{\prime\prime})-{\bf
c_{q-q^\prime}}(t-t^{\prime})\]
\begin{equation}\label{thir}
+\frac c{q_0}\int_t^{t^\prime}dt_1 (t^\prime -t_1)\{{\bf F}[{\bf
r}({\bf q},t_1)]-{\bf F}[{\bf r}({\bf q}^\prime,t_1)]\}.
\end{equation}
The last two terms in Eq. (\ref{thir}) describe the displacement of
two photons from each other because of the difference of their
initial velocities. The term $-{\bf c_{q-q^\prime}}(t-t^{\prime})$
describes the divergence of two straight-line trajectories. The last
term accounts for the different actions of the atmosphere on the
particles moving in different spatial regions.

 Obtaining of the average values in Eq. (\ref{twntei}), which depend
on the wave-vectors  ${\bf r}({\bf q},t^\prime )$ and ${\bf r}({\bf
q^\prime},t^{\prime\prime})$, seems to be challenging because of the
presence of the fluctuating force in ${\bf r}({\bf q},t^\prime )$
and ${\bf r}({\bf q^\prime},t^{\prime\prime})$. Nevertheless the
analysis simplifies if we neglect the correlations between the
forces $F_\alpha$ or $F_\beta$ and the forces entering ${\bf r}({\bf
q},t^\prime )$ or ${\bf r}({\bf q^\prime},t^{\prime\prime})$. This
simplification can be justified by the following reasonings. The
explicit value of the $\alpha$-force is given by
\[F_\alpha[{\bf r(q},t^\prime)]=F_\alpha\big[{\bf r}-{\bf c_q}(t-t^\prime)
-\frac c{q_o}\int_t^{t^\prime}dt_1(t_1-t^\prime)F[{\bf
r(q},t_1)]\big]\]
\begin{equation}\label{tthir}
=F_\alpha\big[{\bf r_\bot}-{\bf c_q}_\bot(t-t^\prime)+c{\bf
e_z}t^\prime -\frac c{q_o}\int_t^{t^\prime}dt_1(t_1-t^\prime){\bf
F[\bf r(q},t_1)]\big],
\end{equation}
where the relation $z=ct$ is used.

If the correlation exists, the distance $|{\bf r}({\bf q},t_1)-{\bf
r}({\bf q},t^\prime)|$ can be estimated by the value
$c(t_1-t^\prime)\leq L_0$. In this case, the integral in Eq.
(\ref{tthir}) is proportional to $(L_0/c)^2$. Hence, the correlation
between $F_\alpha[{\bf r(q},t^\prime)]$ and ${\bf }F[{\bf
r(q},t_1)]$ can be neglected. This approximation implies the
physical picture where the variation of the photon momentum on the
correlation length, $L_0$, is much smaller than $q_0$. Therefore,
the averaging $\langle F_\alpha F_\beta\rangle$ can be performed in
two steps. Firstly, we obtain $\langle F_\alpha F_\beta\rangle$
considering the arguments of $F_\alpha$ and $F_\beta$ to be fixed.
After that, the averaging of the forces, entering the arguments,
should be performed. For example, the term (\ref{twntsi}) is
expressed as
\begin{equation}\label{thir1}
\phi_{PP^\prime}=\omega_0^2\int_0^t\int_0^tdt^\prime
dt^{\prime\prime}\int d{\bf g }\,\psi(g){\bf g}\cdot\big({\bf
p}+{\bf k}t^\prime\frac c{q_0}\big)\,{\bf g}\cdot\big({\bf
p}^\prime+{\bf k}^\prime t^{\prime\prime}\frac c{q_0}\big) \langle
e^{-{i{\bf g}[{\bf r}({\bf q},t^\prime)-{\bf r}({\bf
q}^\prime,t^{\prime\prime})]}}\rangle,
\end{equation}
where the first-step averaging results in appearance of the spectral
density $\psi(g)$. The second-step averaging is shown in
(\ref{thir1}) by the angle brackets. To simplify the derivation of
$\sigma^2$, the authors of \cite{Chu} represented the average of the
exponential function in Eq. (\ref{thir1}) as a product,
\begin{equation}\label{thirrr}
\langle e^{-{i{\bf g}[{\bf r}({\bf q},t^\prime)-{\bf r}({\bf
q}^\prime,t^{\prime\prime})]}}\rangle\approx \langle e^{-{i{\bf
g}{\bf r}({\bf q},t^\prime)}}\rangle \langle e^{i{\bf g}{\bf r}({\bf
q}^\prime,t^{\prime\prime})} \rangle,
\end{equation}
neglecting the correlation of the photon displacements ${\bf r}({\bf
q},t^\prime)$ and ${\bf r}({\bf q}^\prime,t^{\prime\prime})$.
Further analysis explains how this correlation can be accounted for.

First of all, it should be noted that we can integrate Eq.
(\ref{thir1}) over $t^\prime-t^{\prime\prime}$ because of the
presence of the term ${\bf e}_zc(t^\prime-t^{\prime\prime})$ in
${\bf r}({\bf q},t^\prime)-{\bf r}({\bf q}^\prime,t^{\prime\prime})$
[see Eq. (\ref{thir})]. The corresponding fast oscillating function,
$e^{i{\bf e}_z{\bf g}c(t^\prime-t^{\prime\prime})}$, appears in the
last factor of Eq. (\ref{thir1}). Integration of this factor results
in
\begin{equation}\label{thir3}
\int_{-\infty}^{\infty}d(t^\prime-t^{\prime\prime}) e^{i{\bf
e}_z{\bf g}c(t^\prime-t^{\prime\prime})}=\frac{2\pi}c\delta(g_z).
\end{equation}
The lower and the upper limits of the integration over
$t^\prime-t^{\prime\prime}$ are replaced by $\mp \infty$. This can
be approved when the propagation time, $t$, is much greater than
$L_0/c$. In other factors in Eq. (\ref{thir1}), the substitution
$t^{\prime\prime}=t^\prime$ is used.

The relation (\ref{thir3}) means, that only the
$g_{x,y}$-\,components enter Eq. (\ref{thir1}). In particular, the
Fourier-transform $\psi(g)$ should be considered as a function of
the two-dimensional vector ${\bf g}_\perp$: $\psi=\psi\bigg({\sqrt
{g_x^2+g_y^2}}\bigg)$. This observation corresponds to the known
Markov approximation \cite{Tatarskii1} where it is assumed that the
index-of-refraction fluctuations are delta-function correlated in
the direction of propagation. In fact, our derivation, based on the
paraxial approximation, supports the validity of the Markov approach
which at first sight seems to be doubtful.

Using Eqs. (\ref{thir}) and (\ref{thir3}), the expression
(\ref{thir1}) is simplified to
\[\phi_{PP^\prime}=\frac{2\pi\omega_0^2}c\int_0^tdt^\prime \int d{\bf
g }\psi(g)\,{\bf g}\cdot\big({\bf p}+{\bf k}t^\prime\frac
c{q_0}\big)\,{\bf g}\cdot\big({\bf p}^\prime+{\bf k}^\prime
t^{\prime}\frac c{q_0}\big) e^{{i{\bf g}{\bf c}_{\bf
q-q^\prime}(t-t^\prime)}}\]
\begin{equation}\label{thir4}
\times\langle e^{{-i{\bf g}\frac c{q_0}\int_{t^\prime}^t
dt_1(t_1-t^\prime)\{{\bf F}[{\bf r}({\bf q},t_1)]-{\bf F}[{\bf
r}({\bf q}^\prime,t_1)]\}}}\rangle,
\end{equation}
where all the vectors have only the $x-$ and $y$-components, and
${\bf c_{q-q^\prime}}=c({\bf q-q^\prime})/q_0$.

As we see from Eq. (\ref{thir4}), to obtain $\phi_{PP^\prime}$ one
needs to calculate the average value of the exponential function
which is similar to the function in (\ref{twntfo}). Following the
previous procedure, this average can be rewritten as
\[\bigg\langle \exp\bigg\{-{i{\bf g}\frac c{q_0}\int_{t^\prime}^t
dt_1(t_1-t^\prime)\{{\bf F}[{\bf r}({\bf q},t_1)]-{\bf F}[{\bf
r}({\bf q}^\prime,t_1)]\}}\bigg\}\bigg\rangle\]
\begin{equation}\label{thir5}
=\exp\bigg\{-2\pi c^3\int_{t^\prime}^tdt_1(t_1-t^\prime)^2\int d{\bf
g^\prime}\psi(g^\prime)({\bf g}\cdot{\bf g^\prime})^2\big[1-\langle
e^{-i{\bf g^\prime}\cdot[{\bf r}({\bf q},t_1)-{\bf r}({\bf
q}^\prime,t_1)]}\rangle \big] \bigg\}.
\end{equation}
Again, the same function appears in the exponent of the right-hand
side of Eq. (\ref{thir5}) after using the trajectories
(\ref{twntn}). Similar steps can be undertaken many times. In this
way, the time hierarchy, $ 0<t^\prime\leq t_1...\leq t_i\leq t$, is
generated. If the photon-turbulence interaction time, $t-t_i$, is
short, the disturbance of the trajectory is small and vanishes when
$t_i\rightarrow t$. In this case both values, ${\bf r}( {\bf
q},t_i)$ and ${\bf r}( {\bf q}^\prime,t_i)$, approach the value of
${\bf r}$ irrespective of the initial momenta ${\bf q}$ and ${\bf
q}^\prime$. Therefore we substitute the quantity
\begin{equation}\label{thir6}
 \frac12\big\langle\big({\bf g}^\prime\cdot[{\bf r}({\bf
q},t_1)-{\bf r}({\bf q}^\prime,t_1)]\big)^2\big\rangle
\end{equation}
instead of
\begin{equation}\label{thir7}
1-\langle e^{-i{\bf g^\prime}\cdot[{\bf r}({\bf q},t_1)-{\bf r}({\bf
q}^\prime,t_1)]}\rangle
\end{equation}
assuming the exponent in Eq. (\ref{thir7}) to be small. The linear
in ${\bf g^\prime}$ term in the expansion of the exponential factor
is ignored because of its zero-value contribution into the integral
over ${\bf g^\prime}$ in Eq. (\ref{thir5}). Then the term
(\ref{thir7}) reduces to
\[\frac12\big\langle\big({\bf g}^\prime\cdot[{\bf r}({\bf
q},t_1)-{\bf r}({\bf q}^\prime,t_1)]\big)^2\big\rangle\]
\begin{equation}\label{thir8}
\approx\frac {(t-t_1)^2}2\big({\bf c_{q-q^\prime}}\cdot {\bf
g^\prime}\big)^2+\frac{\pi c^3}{30}(t-t_1)^5\int d{\bf
g}^{\prime\prime}\psi(g^{\prime\prime})\big({\bf
c_{q-q^\prime}}\cdot{\bf g}^{\prime\prime}\big)^2({\bf
g}^\prime\cdot{\bf g}^{\prime\prime})^2.
\end{equation}
To obtain Eq. (\ref{thir8}), the approximate relation,
\begin{equation}\label{thir9}
F_\alpha[{\bf r}({\bf q},t_2))]-F_\alpha[{\bf r}({\bf
q}^\prime,t_2))]\approx {\bf c_{q-q^\prime}}(t_2-t)\partial_{\bf
r}F_\alpha[{\bf r}+{\bf c_q}(t_2-t)],
\end{equation}
where $t_1\leq t_2\leq t$, was used. This approximation is in the
spirit of the previous step, where the turbulence effect was assumed
as a small perturbation.

Substitution of Eq. (\ref{thir8}) into the right-hand side of Eq.
(\ref{thir5}) and integration over variables ${\bf g}^\prime$, ${\bf
g}^{\prime\prime}$ and $t_1$ result in
\begin{multline}\label{fort}
\exp\left\{-2.52\cdot10^{-3}C_{n}^{2}{l_0^{\prime}}^{-7/3}c^{3}c^{2}_{{\bf
q}\,{-}\,{\bf q}^{\prime}}(t-t^{\prime})^{5}g^{2}\left[1+\frac{
C_n^{2}{l_0^{\prime}}^{-7/3}c^{3}(t-t^{\prime})^{3}}{560}+\right.\right.\\
\left.\left.+\frac{\cos2\theta}{2}\left(1+\frac{C_n^{2}{l_0^{\prime}}^{-7/3}c^{3}(t-t^{\prime})^
{3}} {2\cdot560}\right)\right]\right\},
\end{multline}
where $l_0^\prime=l_0/2\pi$, and $\theta$ is the angle between the
two-dimensional vectors ${\bf g}$ and ${\bf q}-\,{\bf q}^{\prime}$.

After substitution of (\ref{fort}) into (\ref{thir5}), (\ref{thir5})
into (\ref{thir4}) and (\ref{thir4}) into (\ref{twntsi}), we
calculate $\langle I^2({\bf r}, t)\rangle$. Many-fold integrations
over the variables ${\bf q},{\bf q}^\prime,{\bf p},{\bf
p}^\prime,{\bf k},{\bf k}^\prime,\theta,$ and $t^\prime$ are
performed mainly numerically with employing a computer cluster. In
the course of integration, we have used the Tatarskii modification
of the refractive index spectrum which is derived from the von
Karman form (\ref{add}) by setting $L_0^{-1}=0$. The results for
$\sigma^2$ are shown in Figs. 1-3.

\section{Discussion}

Figs. 1-3 can be used to illustrate the importance of the
correlations of different trajectories.  To simplify our
argumentations, we consider a coherent laser beams, i.e., the case
$r_0=r_1$. Two terms in the last braces of Eq. (\ref{twntt})
contribute equally into $\langle I^2({\bf r},t)\rangle$. Moreover,
if one sets $\phi_{PP^\prime}=0$ in Eq. (\ref{twntfi}), thus
ignoring the correlations of photons with different initial momenta,
we obtain $\langle I^2({\bf r},t)\rangle =2\langle I({\bf
r},t)\rangle ^2$. The scintillation index, $\sigma^2$, is equal to
unity here.
\begin{figure}[h] \center
\includegraphics[height=8cm]{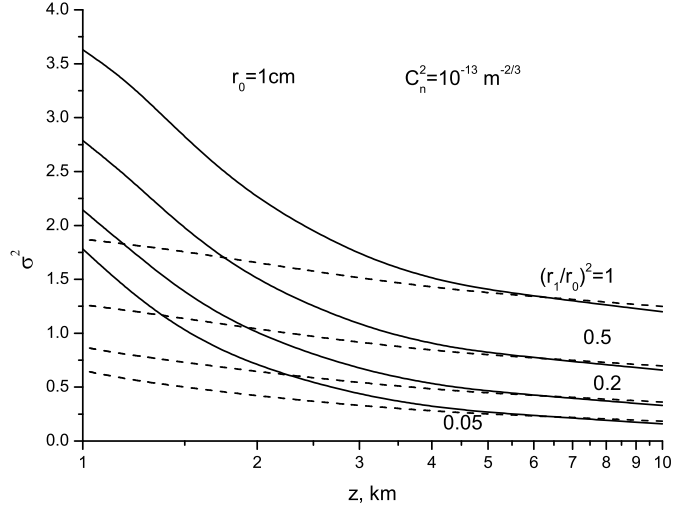}
\caption{Scintillation index of a coherent and partially coherent
beams in the atmosphere versus propagation distance $z$. Dashed
curves correspond to the multiplicative approximation (\ref{thirrr})
for the photon correlations; solid curves are obtained within the
present paper's approach [see Eqs. (\ref{thir5} -\ref{fort})].
$C_n^2=10^{-13}m^{-2/3}$, $r_0 =0.01\,m$, $\frac
{l_0}{2\pi}=10^{-3}m$, and $q_0=10^{7} m^{-1}$. The upper two curves
correspond to the coherent beam.}
\end{figure}
This physical picture is realized for a long-distance propagation
($t\rightarrow\infty$) when the oscillating factor $e^{{i{\bf g}{\bf
c}_{\bf q-q^\prime}(t-t^\prime)}}$ confines the effective volume of
the integration over ${\bf g}$ and ${\bf q-q^\prime}$ to zero [see
Eq. (\ref{thir4})]. For finite values of $t$, the contribution of
$\phi_{PP^\prime}$ becomes quite sizeable that is seen in Figs. 1-3
where the values of $\sigma^2$ are greater than unity.
\begin{figure}%[p]
\center
\includegraphics[height=10cm]{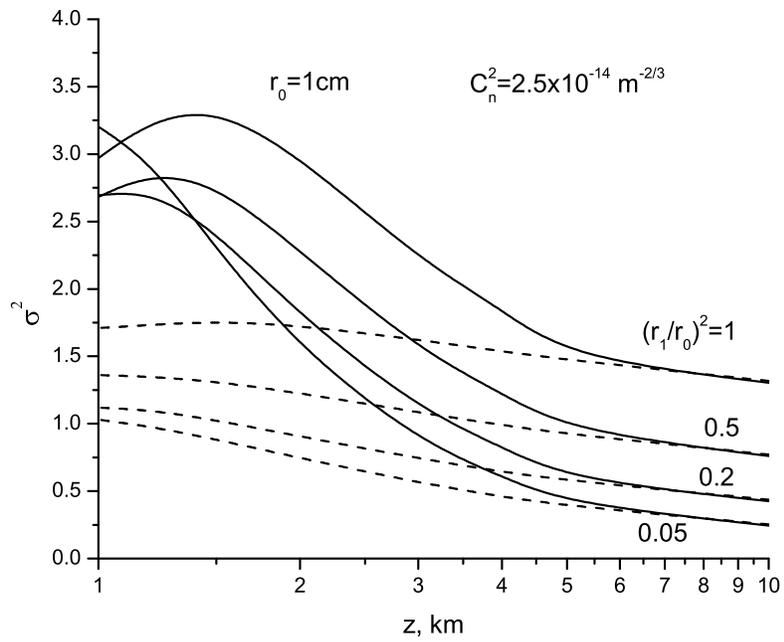}
\caption{The same as in Fig. 1 but for a weaker turbulence strength:
$C_n^2=2.5\times10^{-14}m^{-2/3}$.} \label{three}
\end{figure}
There is a positive contribution of $\phi_{PP^\prime}$ term into the
last exponent in Eq. (\ref{twntfi}) when the vectors ${\bf p,k}$ and
${\bf p^\prime,k^\prime}$ have opposite signs and the difference
$|{\bf q-q^\prime}|$ is not too large. The most favorable conditions
are realized when
\begin{equation}\label{thir66}
 {\bf p}=-{\bf p}^\prime,\quad {\bf k}=-{\bf k}^\prime,\quad {\bf q}={\bf
q}^\prime.
\end{equation}
In this case the sum $\phi_{PP}+2\phi_{PP^\prime}+ \phi_{P^\prime
P^\prime}$ is equal to zero . Eqs. (\ref{thir66}) can be interpreted
as the "super-correlation" conditions under which the value of $M$
is equal to unity and does not depend on the turbulence.

The dependence of $\sigma^2$ on the initial radius $r_0$ can be
explained as follows. The characteristic values of the initial
momentum, ${\tilde q}\sim \sqrt{2}/r_0$, is greater for small $r_0$.
Hence the volume of integration over ${\bf q-q^\prime}$ is also
greater. At the same time the corresponding increase of
$\phi_{PP^\prime}$ occurs only for short distances, $z$, where time
intervals $t$ are sufficiently small and the oscillating factor in
Eq. (\ref{thir4}) is close to unity. Therefore, when $r_0$
decreases, there is an increase of $\sigma^2$ accompanied with the
displacement of the region with enhanced fluctuations towards small
$z$. This is clearly seen in Fig. 3.

In a similar way we can explain a considerable difference of
$\sigma^2$ found  for the plane-wave and spherical-wave models of
radiation in Ref. \cite{andd} (Figs. 1 and 2 there). It follows from
the above reasonings that this effect arises due to very different
initial ${\bf q}$-volumes in the two models.

Also, the  calculations of $\sigma^2$ in the Ref. \cite{Ban54}
should be mentioned where a simplified model of the turbulence was
used (see Fig. 1 there). The results of Ref. \cite{Ban54} well
correlate with ours.

Comparing the results of the present paper and those, based on the
approximation of uncorrelated trajectories (\ref{thirrr})
(respectively, solid and dashed lines in Figs. 1 and 2), we see  a
more pronounced growth of $\sigma^2$ at a moderate turbulence in the
former case. Figures 1-3 illustrate that this holds true for the
distances of $1-3$ km. We attribute the evident distinction of the
results to a better accuracy of accounting for the correlations of
the photon trajectories. At the same time, both approaches provide
the known in the literature saturation effect: $\sigma^2\rightarrow
r_1^2/r_0^2$ when $z\rightarrow\infty$.

The phase diffuser with a short characteristic time (a
high-frequency diffuser) does not change qualitatively the physical
picture described above. At the same time, both approaches reveal an
ability of the diffuser to suppress scintillations which
 is favorable for communication performances.

The effect of the phase diffuser is explained as follows. The
initial phase relief, introduced by the diffuser, varies in time.
The photon trajectories depend on the initial state of the radiation
and varies synchronously with the diffuser state. A ``slow" detector
integrates the contribution of these photons. Although the
atmosphere stays almost frozen during the integration time, the
diffuser provides a better averaging of the propagating radiation
over the refractive-index relief. Therefore, the fluctuations of the
detected signal decrease.

This is not a unique way to suppress fluctuations. For example, the
authors of Ref. \cite{gorsh} proposed to use asymmetric optical
vortices. The range of a weak and moderate turbulence was studied.
Numerical simulations of the beam propagation  showed promising
results. It should be emphasized that in this case the experimental
setup does not require a high-frequency phase diffuser.

\begin{figure}%[p]
\center
\includegraphics[height=10cm]{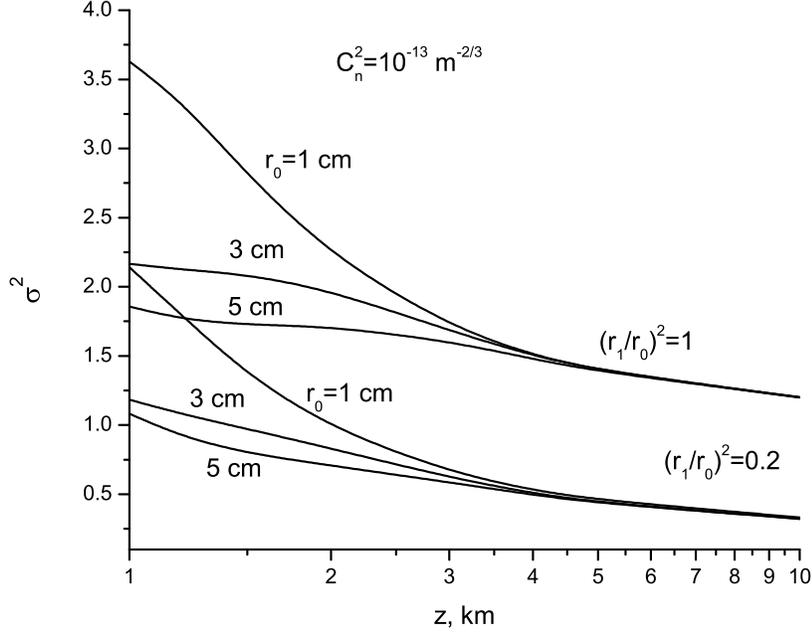}
\caption{Scintillation index versus propagation distance $z$ for
different initial radii of the beam: $r_0=0.01 m$, $0.03 m$, $0.05
m$. The rest of the parameters are the same as in Fig 1.}
\label{one}
\end{figure}

\section {Applicability of the distribution function approach for
short distances}

Our analysis is based on Eq. (\ref{twntt}) obtained within the
concept of photon trajectories. To consider photons as particles,
whose density in the $({\bf r},{\bf q})$ domain is defined by the
distribution function $f({\bf r},{\bf q},t)$, the uncertainty of the
momentum, ${\bf q}$, should be small. The value of the uncertainty
can be estimated from the definition of the distribution function
(\ref{threee}) as ${\tilde k}/2$. It follows from Eq.
(\ref{sevntn}{}) that close to the source and in the absence of the
diffuser the ratio $\frac{{\tilde q}}{{\tilde k}/2}\sim\frac{\langle
q^2\rangle^{1/2}}{\langle{k^2/4} \rangle^{1/2}}=
\frac{(2/r_0^2)^{1/2}}{(2/r_0^2)^{1/2}}=1$. Hence in the vicinity of
the source, our calculations of $\sigma^2$ are not applicable if the
light is in  a coherent state.

The situation changes drastically for a remote detector. With
increase of the propagation path, z, the value of ${\tilde q}$
increases.  The corresponding gain of the photon momentum, $\Delta
{\bf q}$, is generated by a random force, ${\bf F}$. Hence the
average value, $\langle{\Delta\bf q}\rangle$, is equal to zero while
the nonzero mean-square value is given by \cite{Chu}
\begin{equation}\label{thir67}
\langle\Delta
q^2\rangle=0.066\pi^2\Gamma(1/6)q_0^2{l_0^\prime}^{-1/3}C_n^2z.
\end{equation}

In contrast to ${\tilde q}$, the value of ${\tilde k}$ decreases
because of the broadening of the beam. The mean-square of the beam
radius is given by \cite{Fante1,Chu}
\begin{equation}\label{thir68}
R^2=
\frac{r_0^2}2\big[1+\frac{4z^2}{q_0^2r_0^2r_1^2}+\frac{8z^3T}{r_0^2}\big],
\end{equation}
where $T=0.558l_0^{-1/3}C_n^2$. When the last term in square
brackets  dominates, the ratio ${\tilde q}^2/({\tilde k}/2)^2$ can
be estimated as
\begin{equation}\label{thir69}
\langle\Delta q^2\rangle R^2\approx 15\cdot q_0^2l_0^{-2/3}C_n^4
z^4,
\end{equation}
where $\langle\Delta q^2\rangle$ is assumed to be of the order of
${\tilde q}^2$ thus ignoring the square of the initial momentum
$2/r_0^2$.

Substituting $z=10^3 m$, $q_0=10^7 m^{-1}$, and
$l_0=2\pi\cdot10^{-3} m$ into Eq. (\ref{thir69}), we obtain $\frac
{\tilde q}{{\tilde k}/2}\sim 21$ that provides adequacy of our
approach for the whole range of $z$ variations shown in Figs. 1-3.
This range concerns not only coherent, but also partially coherent
beams. For partially coherent beams, the minimum $z$ can be even
smaller than for coherent beams. This is because of an additional
diffuser-caused growth of ${\tilde q}^2$, which is estimated by the
value ${\Delta\tilde q}^2_{diffuser}\sim 2/r_1^2$.

\section{Conclusion}
The paper continues the studies presented in Refs. \cite{Chumak
wander,Chu}. Using the approach of the distribution function, the
problem of obtaining of $\sigma^2$ reduces to calculation of the
correlations between different photon trajectories. Assuming the
outer radius of turbulent eddies much smaller than the propagation
distance, the iterative procedure for calculations of these
correlations is developed. The modified approach makes it possible
to extend  applicability of the theory to a wider range of the
propagation distances. This range includes a strong turbulence as
well as a considerable part of a moderate turbulence where the
scintillation index tends to reach its maximum value. The criterium,
derived in Sec. 5, imposes the restriction on our theory from the
side of short distances (weak turbulences).

\section{Acknowledgments}
The authors thank V. Bondarenko, V. Gorshkov and A. Semenov for
useful discussions and comments.


\begin{thebibliography}{99}


\bibitem{Schuster}A.P. Schuster, Astrophys. J., {\bf 21} (1905).

\bibitem{mil} P.W. Milonni, J.H. Carter, J.C. Peterson, and R.J. Hughes,
J. Opt. B: Quantum Semiclass. Opt. {\bf 6}, S742 (2004).

\bibitem{Tatarskii1} V.I. Tatarskii, The effect of the Turbulent Atmosphere on Wave
Propagation. Springfield, VA: National Technical Information
Service, U.S. Department of Commerce, (1971).

\bibitem{Andrews} L.C. Andrews and R.L. Phillips, Laser Beam Propagation Through Random Media.
Bellingham, WA: SPIE Press (1998).

\bibitem{Andrews1} L.C. Andrews, R.L. Phillips, and C.Y. Hopen, Laser Beam
Scintillation with Applications. Bellingham, WA: SPIE Press (2001).

\bibitem{Fante1} R.L. Fante, Proc. IEEE, {\bf 63} (1975).

\bibitem{xia} X. Liu, F. Wang,and Y. Cai, Opt. Lett., {\bf 39}, 3336, (2014).

\bibitem{Chumak wander} G.P. Berman, A.A. Chumak, and V.N. Gorshkov, Phys. Rev.E {\bf 76},
056606 (2007).

\bibitem{gra} M.E. Gracheva and A.S. Gurvich, Izv. Vysch. Uchebn. Zaved.
Radiofiz., {\bf 8}, 717 (1965); M.E. Gracheva, A.S. Gurvich, and
M.A. Kallistratova, Radiophys. Quantum Electron, {\bf 13}, 40
(1970).

\bibitem{fan} R.L. Fante, Proc. IEEE, {\bf 68}, 1424 (1980).

\bibitem{Wang} S. Wang, M. Plonus, and C. Ouyang, Appl. Optics, {\bf
18}, 1133 (1979).

\bibitem{fan2} R.L. Fante, IEEE Tranc. Antennas Propagat., {\bf AP-25}, 266 (1977).

\bibitem{Lee} M. Lee, J. Holmes, and J. Kerr, J. Opt. Soc. Amer., {\bf 67}, 1279 (1977).

\bibitem{Ban55}  V.A. Banakh and V.M. Buldakov,
\newblock Opt. Spectrosk., {\bf 55}, 707 (1983).

\bibitem{Ban54}  V.A. Banakh, V.M. Buldakov, and V.L. Mironov,
Opt. Spectrosk., {\bf 54},  1054 (1983).

\bibitem{Chu} G.P. Berman and A.A. Chumak, Phys. Rev.
A,{\bf 74}, 013805 (2006).

\bibitem{cor} O. Korotkova, L.C. Andrews, and R.L Phillips, Proceedings of SPIE,
 {\bf 4821}, 98 (2002).

\bibitem{coro} O. Korotkova, L.C. Andrews, and R.L Phillips, Opt.Eng.,
 {\bf 43}, 330 (2004).

\bibitem{fei} F. Wang, X. Liu, and Y. Cai, Prog. in Electromagn.
Res., {\bf 150}, 123 (2015).

\bibitem{Das} R. Dashen, J. Math. Phys., {\bf 20}, 894 (1979).

\bibitem{Yak} I.G. Yakushkin, Radiophys. Quantum Electron., {\bf 19}, 270 (1976).

\bibitem{Fed} A. Fedrizzi, R. Ursin, T. Herbst, M. Nespoli1, R. Prevedel, Th.
Scheidl, F. Tiefenbacher., Th. Jennewein, and A. Zeilinger, Nature
Phys. {\bf 5}, 389 (2009).

\bibitem{Hug} R. Hughes, J. Nordholt, D Derkacs and Ch. Peterson, New J.
Phys. {\bf 4}, 43.1 (2002).

\bibitem{Bou} J. Bourgoin, E. Meyer-Scott, B. L. Higgins, B. Helou,
C. Erven, H. Hubel, B. Kumar, D. Hudson,I. D'Souza, R. Girard, R.
Laflamme, and T. Jennewein, New J.Phys, {\bf 15}, 023006 (2013).

\bibitem{Sem} A. Semenov and W. Vogel, Phys Rev. A  {\bf 81}, 023835 (2010);
 D. Vasylyev, A. Semenov, and W. Vogel,  Phys Rev. Lett., {\bf 108},
220501 (2012); V. Usenko et al., New J. Phys, {\bf 14}, 093048
(2012).

\bibitem{UJP} A.I. Rarenko,  A.A. Tarasenko, and A.A. Chumak, Ukr. J. Phys., {\bf
37}, 1577 (1992); O. Chumak and N. Sushkova, Ukr. J. Phys., {\bf
57}, 30 (2012).

\bibitem{ChuSingle} G.P. Berman and A.A. Chumak, Proc. of SPIE, {\bf
6710} (2007).

\bibitem{ChuPhase} G.P. Berman and A.A. Chumak, Phys. Rev. A,{\bf
79}, 063848 (2009).

\bibitem{sto} O.O. Chumak and E.V. Stolyarov, Phys. Rev. A, {\bf 88}, 013855 (2013).

\bibitem{sto14} O.O. Chumak and E.V. Stolyarov, Phys. Rev. A, {\bf 90}, 063832 (2014).

\bibitem{chuZ} A. A. Tarasenko and A. A. Chumak, JETP {\bf 73}, 625 (1977)

\bibitem{stroh} J. Strohbehn and S. Clifford, IEEE Trans. Antennas
Propag., {\bf AP-15}, 416 (1967).

\bibitem{krav} Yu.A. Kravtsov, Rep. Prog. Phys.,{bf 55}, 39(1992).

\bibitem{andd} L.C. Andrews and R.L. Phillips, SPIE, {\bf 3609}, 90
(1999).

\bibitem{gorsh} G.P. Berman, V.N. Gorshkov, and S.V. Torous,
J. Phys. B: At. Mol. Opt. Phys. {\bf 44}, 055402 (2011).

\end{thebibliography}
\end{document}